# NOVEL HYBRID INTRUSION DETECTION SYSTEM FOR CLUSTERED WIRELESS SENSOR NETWORK

Hichem Sedjelmaci[1] and Mohamed Feham[1]

STIC Lab, Department of telecommunications, Abou Bakr Belkaid University, Tlemcen, Algeria
`massy2008@hotmail.fr, m_feham@mail.univ-tlemcen.dz`

## ABSTRACT

*Wireless sensor network (WSN) is regularly deployed in unattended and hostile environments. The WSN is vulnerable to security threats and susceptible to physical capture. Thus, it is necessary to use effective mechanisms to protect the network. It is widely known, that the intrusion detection is one of the most efficient security mechanisms to protect the network against malicious attacks or unauthorized access. In this paper, we propose a hybrid intrusion detection system for clustered WSN. Our intrusion framework uses a combination between the Anomaly Detection based on support vector machine (SVM) and the Misuse Detection. Experiments results show that most of routing attacks can be detected with low false alarm.*

## KEYWORDS

*Wireless Sensor Network, Hybrid Intrusion Detection System, Support Vector Machine (SVM), Classification Accuracy, False alarm*

## 1. INTRODUCTION

Recent advances that deal with the technology of micro-electronics and wireless communication have enabled the development of multifunctional sensor with low-cost and low-power. These sensor nodes consist of data processing, wireless communication and capture device.

A wireless sensor network (WSN) consists of a large number of devices operating independently and communicating with each other via short-range radio transmissions. These sensors can be very useful for many military and civilian applications, as collecting and processing information from hostile environment and difficult access locations such as battlefield surveillance, environment monitoring, etc. Many researchers have focused on the security of WSN against attacks or malicious behaviors since that the characteristics of both wireless infrastructure and WSNs can cause the potential risk of attacks on the network [1]. The security mechanisms used to protect the wireless sensor network against intruders are:

Cryptographic techniques: They are used to ensure authentication and data integrity by checking the source of the data and to verify that is not altered. The cryptographic operations are based on primitives such as hash functions, symmetric encryption and public key cryptography [2] which can protect WSN against external attacks. However cryptographic techniques cannot detect internal attacks when the attacker knows the keys and uses them to perform encryption/decryption. This technique is defined as the first line of defence.

Steganography: If cryptography is the art of secrecy, steganography is the art of concealment. The main objective of steganography is hiding or embedding a message either in another one or into a multimedia data (image, sound, etc). However this technique requires significant processing resources and it is hard to implement it in WSN because of the constraints of these sensors.

IDS: An intrusion detection system is usually considered as the second line of defence, it can protect with high accuracy against internal attacks. This mechanism allows detecting abnormal or suspicious activities on the analyzed target and triggers an alarm when intrusion occurs. As far as cryptography is concerned it cannot provide the necessary security in WSN, that why we believe strongly that IDS is useful for both internal and external attacks. Many researches in the application of the IDS' technology in ad hoc networks were done, in comparison with wireless sensor networks where few subjects were investigated because of its limited energy and computing storage capacity.

Our aim is to introduce in the following research a novel hybrid intrusion detection system for WSN. The approach uses the clustering algorithm to reduce the amount of information and decrease the consumption of energy. In addition we have used a class of machine learning algorithm called support vector machine (SVM), that separates data into normal and anomalous (binary classification), in order to detect anomalies. We have also applied misuse detection technique to determine known attack patterns (signatures). Therefore, the combination of both techniques can achieve high detection rate with low false positive and false negative rate. Finally, we have developed a mechanism of cooperation among IDS agents that work with each other, that mechanism can make a better decision in order to verify if a node is compromised or not which might determine novel sign of intrusion.

## 2. RELATED WORK

Sensor networks introduce severe resource constraints due to their lack of data storage and power [3], according to Roman et al. [4] IDS solutions for ad hoc networks cannot be applied directly to sensor networks. Therefore, the proposed intrusion detection system must meet the demands and restriction of WSNs.

Kumar [5] classified intrusions detection techniques into two categories:

• Misuse detection: this technique involves the comparison between captured data and known attack signatures, any corresponding pattern can be considered as an intrusion. Updating the signature over time is necessary to keep this technique effective. However, the major drawbacks of misuse detection systems are their inability to detect unpublished attacks [6].

• Anomaly detection: is based on modeling the normal behavior of the nodes and compares the captured data with this model, any activities that deviate from this model can be seen as an anomaly. The advantage of such technique is that it can detect attacks that are unpublished [6]. On the contrary this technique requires a considerable computation time which implies high energy consumption. Therefore, the anomaly detection algorithm in WSN must take into consideration a Trade-off between detection accuracy and energy consumption. Among anomaly detection techniques proposed in literature for wireless sensor networks are: Rule Based, K-Nearest Neighbor and support vector machine (SVM), etc [7].

### 2.1 Anomaly detection based on SVM

There have been currently limited researches on the use of SVM classifier in WSN. Kaplantzis et al. [8] worked on centralized intrusion detection system based on support vector machine to detect selective forwarding and black hole attacks. IDS that are run in the base station use one-class SVM in training collected nodes' data for only normal network activity, by using bandwidth and hops count as feature vector (no attacks activity). Subsequently, the attacks are introduced in the network, and SVM use the training set to detect the attacks. In the simulations, the authors claim that the IDS can detect with high accuracy black hole and selective forwarding attacks. However, this scheme can detect only two kinds of attacks, and presents low detection rate for the selective forwarding attack when its number is small in the network. Besides, the base station cannot manage the large number of packets sent by the nodes which consequently cannot be analyzed by the SVM.

Centralized SVM training method allows a better separation of the classes with the misclassification error rate close to zero [9]. However, it exhibits a high communication overhead, and it is not suitable for resource-constrained sensor networks. That is why many authors mentioned that the distributed SVM training fit the requirement of sensor nodes in terms of energy cost ([9], [10], [11], [12]) and provide closer classification as centralized approach. In [12], two distributed algorithms for training SVM in WSN are proposed. For both algorithms the SVM classifiers are run in each node to compute a set of data vector. For the first algorithm, it is called Support Vectors (that lie closet to the separating lines), but for the second one the amount of information is much larger, it represents the vectors that lie in the convex hulls boundary of each of the two classes (normal, anomalous). Each node communicates its data vector with one-hop neighbor, once this process done, the final hyperplane is computed, and all the nodes have the same discriminant plan to separate data into two classes and can classifier any new measures. In the simulations the second algorithm exhibits a better classification than the first algorithm but with additional power consumption.

## 2.2 Hybrid intrusion detection system integrated for clustered sensor networks

There are some researches that use a combination between anomaly detection and misuse detection (hybrid model) in order to leverage the advantages of these two techniques and try to detect a significant number of attacks. In the literature there are some hybrid intrusion detection systems for WSN such as [13], [14] and [15].

In [13], Hai et al. proposed in WSN cluster based and hybrid approach for IDS. Based on work undertaken by Roman et al. [4], they suggest that IDS agents are located in every node. The agent is divided into two modules: local IDS agent and global IDS agent. Because of energy and memory constraints of WSN, global agents are active only at a subset of nodes. For anomaly detection, the global agent IDS monitors the communication of its neighbors by using predefined rules with two-hop neighbor knowledge, then sends alarm to cluster-head (CH) when they detect malicious nodes. Each node has an internal malicious database, which contains a list of known attack patterns (signatures) computed and generated in the CH.

The authors attempt to minimize the number of nodes where the global agents IDSs are deployed by evaluating their trustworthy based on trust priority. In order to reduce the collisions and avoid the waste of energy, they propose an over-hearing mechanism that reduces the sending message alerts. When the rate of collision and the number of malicious node is not very high the proposed scheme can detect the routing attacks such as selective forwarding, sinkhole, hello flood and wormhole attacks with a better energy saving. Nevertheless, the drawback of this scheme is the high rate of false positive that is generated when using the rule based-approach of anomaly detection. In addition, this method is well defined by experts and specialists in the area of wireless security by being dependent on manual rule updating.

Yan et al. [14] have focused on using clustering approach in WSN and embedded hybrid IDS in CH. the proposed IDS have three modules: misuse detection module, anomaly detection module and decision making module.

In the anomaly detection module, the rule-based method has been used to analyze incoming packets and categorize the packet as normal or abnormal. For building misuse detection model, the supervised learning algorithm Back Propagation Network (BPN) is adopted. The abnormal packets, which are detected by anomaly detection model is used as input vectors of BPN. The algorithm trains this training dataset, then classifies the data into five classes (four types of attacks and one normal behavior), when the process of training is over, it integrates the model in the misuse detection module in order to classify the new data (testing dataset).

Finally the output of both models (anomaly detection and misuse detection models) is used as an input for the decision making module. The rule-based method is applied to determine if an incoming information is an intrusion or not, and determine the category of attack. In case of presence of an intrusion the module reports the results to the base station. The simulation results show a higher rate of detection and a lower false positive rate, but the major drawback of the proposed scheme is that IDS monitor run in a fixed cluster heads (the hot point). Therefore it's an attractive node for the intruder that uses all its capacity to attack this node. Another drawback is the number of features which is very important (twenty four features are used). Thus the cluster head consumes much more energy, which leads to minimize the life time of the node, also the names of this features are not mentioned.

## 3. ROUTING ATTACKS IN WSNS

A large variety of attacks against WSNs exist in the literature; therefore in this section we specify two categories of attacks: Dos attack and Probe attack.

Select Forwarding and black holes use illegitimate data forwarding to make attack, so they are classified as Dos attack [14].

Spoofed, Altered or Replayed Routing Information, Wormholes and Acknowledgment Spoofing make a probe step before they begin to attack, so they are classified as Probe attack [14].

• Selective forwarding: In a selective forwarding attack, the intruders prevent the forwarding of certain packets by dropping them. They can also forward a received packet along a wrong path, thus creating unfaithful routing information in the network [1].

• Black holes: In this attack, the intruder pretends to be as shortest path to the base station or cluster head (CH) by using a higher power transmission. The WSN are vulnerable of this kind of attacks because of their communication paradigm, where all nodes carry data to the single node, in our case, the CH.

• Wormholes: The attacker tunnels packets received at one location in the network (in our case, the cluster) to another location, where the packets are then replayed. khalil et al. propose five modes of wormhole attacks in WSN, the detailed of these modes are defined in [16].

• Spoofed, Altered or Replayed Routing Information: The attacker monitors transmissions, intercepts packets, then altering or repelling traffic, this attack can also lead to create routing loops in networks [6].

Acknowledgment Spoofing: In this attack, the intruder convincing the sender that a weak link is strong or that a dead node is alive [6]. This result in packets sent along such link or node are lost.

## 4. SVM FOR ANOMALY DETECTION

In this section a brief description of SVMs and feature selection are presented

### 4.1 Support vector machines

Support vector machines are a set of supervised learning techniques used for regression and classification .The aim of SVM classifier is to determine a set of vector called support vector to construct a hyperplan in the feature spaces.

There are several researches that use SVM based on multi-class for traditional network to classify a data into n-classes, but this approach don't meet the requirement of sensors network and remains as an open research question. In our context a distributed binary classifier (anomaly or normal) for anomaly detection is performed to detect the abnormal packet.

Given the training datasets , $(x_i, y_i)\ i = 1, \ldots n,\ y_i \in \{-1, +1\}, x_i \in R^d$ , we want to find the hyperplane that have a maximum margin:

$$w.x = b$$

Where $w$ is a normal vector and the parameter b is offset.

In order to find the optimal hyperplane, we must solve the following convex optimization problem:

$$\left. \begin{aligned} & \min \left\{ \frac{\|w\|^2}{2} + C \sum_{i=1}^{n} \varepsilon_i \right\} \\ & y_i(w.x_i + b) \geq 1 - \varepsilon_i\ ,\varepsilon_i \geq 0, 1 \leq i \leq n \end{aligned} \right\} \quad (1)$$

$\sum_{i}^{n} \varepsilon_i$ relax the constraints on the learning vectors, and C is a constant that controls the tradeoff between number of misclassifications and the margin maximization.

The Eq. (1) can be deal by using the Lagrange multiplier [17]:

$$\left. \begin{aligned} & \text{maximize } L(\alpha) = \sum_{i=1}^{n} \alpha_i - \frac{1}{2} \sum_{i=1}^{n} \sum_{j=1}^{n} \alpha_i \alpha_j y_i y_j K(x_j, x_i) \\ & \text{subject to } \sum_{i=1}^{n} y_i \alpha_i = 0, \text{and } 0 \leq \alpha_i \leq C \text{ for all } 1 \leq i \leq n \end{aligned} \right\} \quad (2)$$

Here $K(x_j, x_i)$ is the kernel function and $\alpha_i$ are the Lagrange multipliers. According to the condition of Kuhn-Tucker (KKT), the $x_i$s that corresponding to $\alpha_i > 0$ are called support vectors (SVs).

Once the solution to Eq. (2) is found, we can get [17]:

$$w = \sum_{i=1}^{n} \alpha_i y_i x_i \quad (3)$$

Thus the decision function can be written as:

$$f(x, \alpha, b) = \{\pm 1\} = sgn\left( \sum_{i=1}^{n} y_i \alpha_i k(x, x_i) + b \right) \quad (4)$$

We choose SVM classifier for anomaly detection because it's provide very good results with less training time compared to neural networks. In addition, it is more suitable for intrusion detection in case where new signature is detected. Another advantage of SVM is the low expected probability of generalization errors [18].

### 4.2 Feature selection

Feature selection is an important factor to increase the classification accuracy, reduce the false positive and get a fast training time. In this research, the feature selection method proposed by Sung et al. [18] is adopted. Thus the most relevant features are selected.

## 5. PROPOSED FRAMEWORK AND ITS WORKING

The novelty of our approach is using hybridization between anomaly detection based on SVM and misuse detection, in order to achieve a more accurate intrusion detection system. The anomaly detection uses a distributed learning algorithm for the training of a SVM to solve the two-class problem (distinguish between normal and anomalous activities). In addition, we use a hierarchical topology that divide the sensor network into clusters, each one having a cluster head (CH). The objective of this architecture is to save the energy that allows the network life time prolongation. Among the Cluster-based routing protocols founded in the literature are: LEACH [19], PEGASIS [20], HEED [21]. At last, each node has the possibility to activate its IDS. Activating every node as an IDS wastes energy. So minimization of number of nodes to run intrusion detection is necessary [22]. We defined N as the average number of IDS nodes for each individual link that is expressed by the following equation [16]:

$$N = 1.6 r^2 d$$

Where d is network density and r is the communication range.

Each IDS monitors the neighbor nodes with no trust between each pair of agents (i.e. IDS also monitor its IDS neighbor) as illustrated in Figure 1.

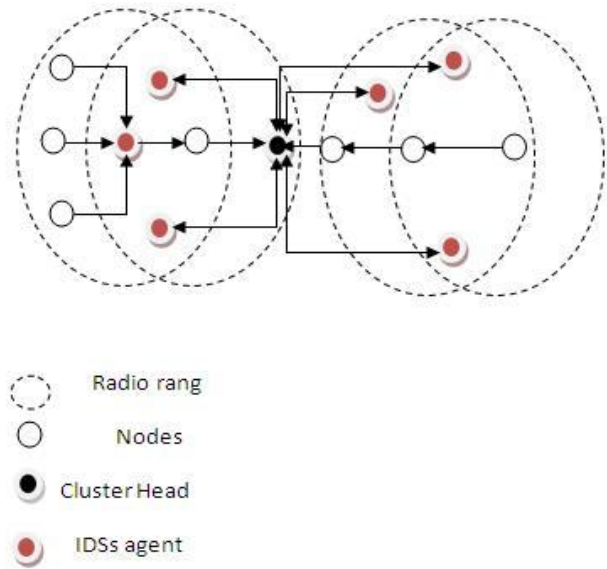

Figure 1. Placement strategy of IDSs node illustration

In our model, we assume that the sensor nodes are stationary and cluster head has more energy compared to the other ones. In the training phase, each IDS node receives the data (support vector) from the nearby IDS nodes by keeping its radio in a promiscuous mode or through multi-hop communication mode (the cluster head can act as a relay). We also make the assumption that the communication activity in this case is supposed to be secured (the algorithm is detailed bellow). In the end, we embed the selected training model into hybrid intrusion detection module (HIDMs) in order to obtain a lightweight and accurate detection system. The selected model is chosen according to the processes illustrated in Figure 2.

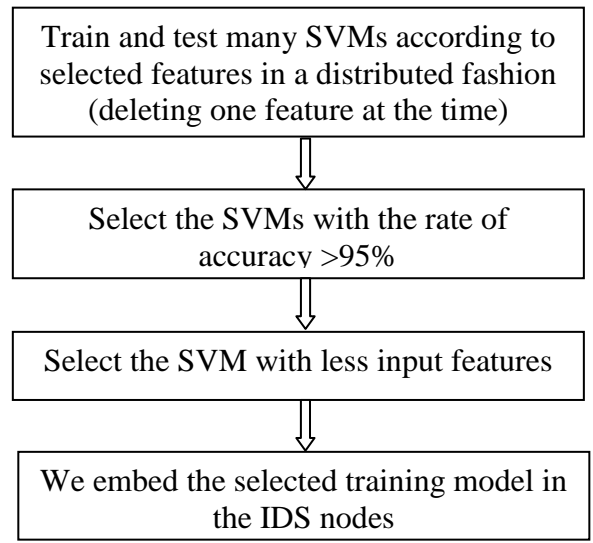

Figure 2.Optimal distributed SVM selection process

## 5.1 The IDS agent architecture

The proposed intrusion detection system (Figure 4) comprises three modules, which are detailed as follow:

### 5.1.1 Data Collection Module (DCM)

Due to broadcast nature of wireless networks, monitor nodes gather the packets within their radio range [13] and pass it to the Hybrid Intrusion Detection Module.

### 5.1.2 Hybrid Intrusion Detection Module (HIDM)

The hybrid intrusion detection module involves anomaly and misuse detection techniques.

   A.  SVM-Based Anomaly Detection Engine

The anomaly detection procedure is divided into two stages:

**Stage1: The training process.** Each IDS agent trains the SVM locally, then computes a set of data vectors called support vector (these set of data vectors are less in number than the input data vector used during the learning process). These later will be sent to an adjacent IDS node that is situated in the same cluster.  Each monitor node that receives support vector from their IDS neighbors or cluster Head makes a combination between the union of received set and its own support vectors. These monitors update their support vector and compute the separating hyperplane. Afterward, they transmit the resulted set of support vector to the nearby IDS nodes. This process is continued until all IDS agents in the same cluster reach the same trained SVM (a complete pass through all IDSs within the same cluster). The communication activity within the cluster between IDS nodes are depicted in Figure 3. For each cluster, the selected IDS agent that depends on its residual energy, sends its support vector to the concerned cluster head; then, all the cluster heads exchange their data and communicate the computed set of support vector to their IDS nodes. Finally, when they all compute the global vector support the result is the same, after that, they can classify new captured packets as normal or anomalous. This algorithm reveals little communication overhead and less power consumption since the communication is performed only with a vector support rather than the whole data as in the case of the centralized approach. In addition, in order to save the energy, each IDS node sends back different values of support vector from the ones sent before.

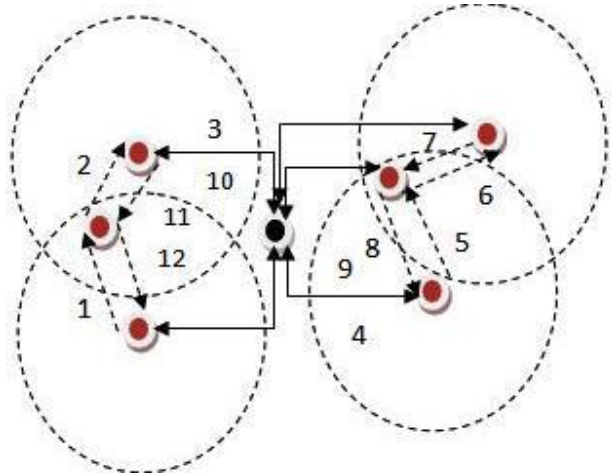

- - -► : Communication performed in a promiscuous mode
⎯⎯► : Communication performed through multi-hop relay.

Figure 3.Communication of support vector between IDS nodes

**Stage2: SVM testing process.** When the process of training is over each IDS node classifies the new data according to normal and anomaly patterns.

The selected training model (described above) is used for anomaly detection engine to classify the captured data that are delivered from data collection module. Any deviations from normal pattern are considered as an intrusion and delivered to misuse detection engine for further detection.

### B. Signature Based Detection Engine:

When misuse detection engine receives the intrusion report (the suspected node, a set of features) from anomaly detection engine, it uses some predefined signs of intrusion that are stored in the signature database to check the occurrence of intrusion. If match occurs, the IDS node sends an alarm to cluster head that the analyzed node is an intruder. The cluster head removes the compromised node from the cluster and inform its IDS agents and all CHs over the network about the malicious node. If no match occurs, the process of cooperation is launched. Note that we stored at all nodes in the network a predefined rule about a set of intrusion signature.

### 5.1.3 Cooperative Detection Module (CDM)

If there are no matches between the intrusion detected by anomaly detection engine and some predefined signatures of attacker, the IDS agent sends the intrusion report to cluster head. That node performs a voting mechanism to make a better decision about the suspect nodes.  If more than half of IDS nodes within the same cluster claim that the analyzed target is an attacker, the cluster head isolates the suspect nodes from the cluster and compute a new rule regarding the novel intrusion, then sends an alert message (that include a malicious node and novel intrusion signature) to their IDS agents and all CHs over the network. When the IDS agents receive this message they update their signature database.

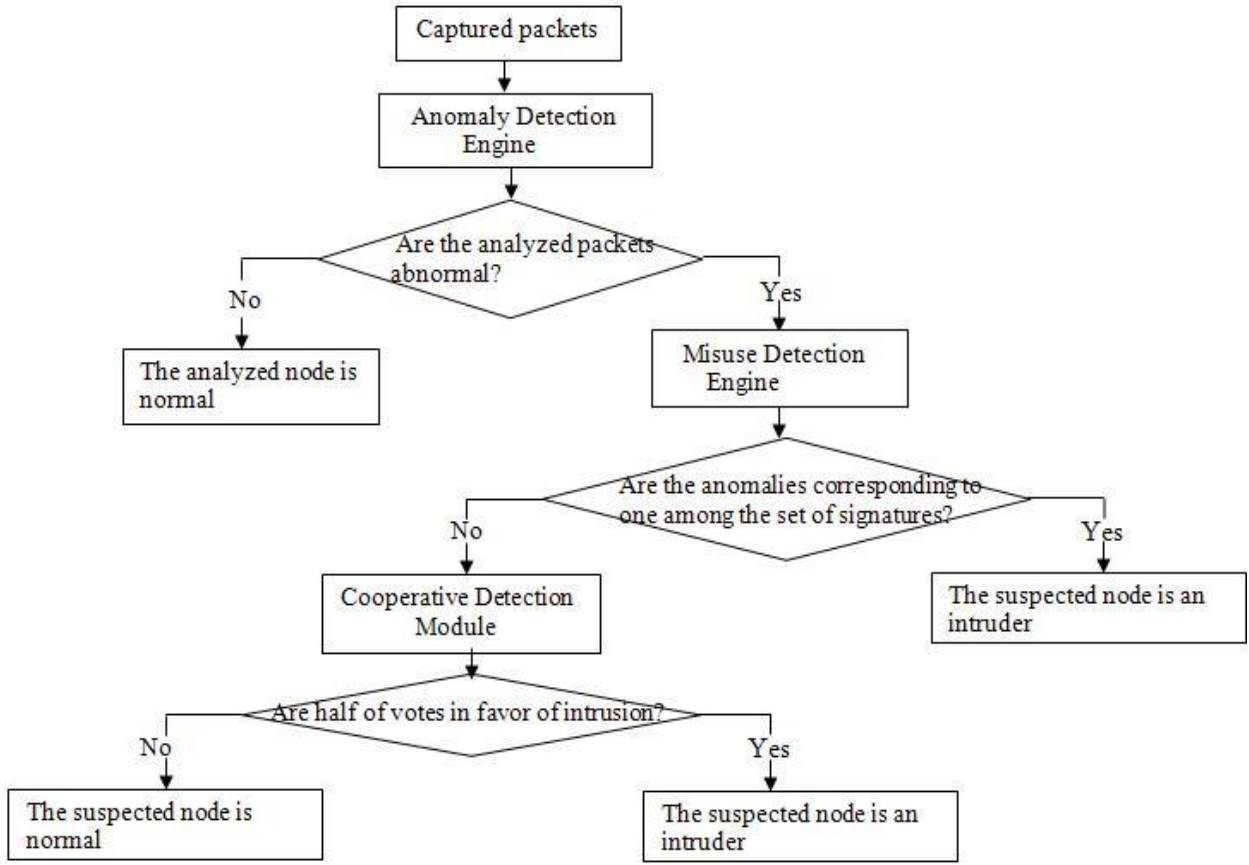

Figure 4. The flowchart of IDS framework in WSN

### 5.2 Dynamic process for intrusion detection system

In the suggested approach, if (3/4) of IDS nodes within the cluster have consumed more than 50% of their energy; new IDSs are elected and receive the actual set of intrusion signature from the cluster head. The older ones are designated as ordinary nodes. Note that the new IDSs election depends on the residual energy and the placement strategies suggested by Khalil et al. [16]. This mechanism helps to avoid depletion of energy nodes, thus prolonging the network lifetime. When new IDS nodes are elected, they compute locally the support vector and the distributed algorithm for training SVMs is performed as alluded above.

## 6. EXPERIMENTS

In this section we evaluate the performance of the proposed hybrid IDSs. In our experiments, we have used the KDDcup'99 dataset [23] as the sample to verify the efficient of the distributed anomaly detection algorithm and valid it by compare with a centralized SVM-based classifier, which achieve a high level of accuracy detection. Also, we compare the distributed hybrid IDSs with one proposed by Yan et al. [14] and Hai et al. [13] according to the false positive rate (false alarm) in order to determine the effectiveness of our scheme.

## 6.1 Dataset

The KDD 99 intrusion detection dataset is developed by MIT Lincoln Lab in1998, each connection in the dataset has 41 features and it's categorized into five classes: normal and four attack behaviors (Dos, Probe, U2r, R2l).

Our analysis is performed on the "10% KDD" intrusion detection benchmark by using its samples as training and testing dataset. We focus only on two categories of attacks (Dos and Probe attacks), which are defined as anomalies behavior and are classified as (-1). The normal behavior is classified as (+1).

The training data used at each IDS comprises of 50 normal and 50 anomalous samples (include both Dos and Probe attacks). In order to evaluate the proposed algorithm, the amount of the data used in test process is equal to N*60, where N is the number of IDS nodes in the network, and the amount of both anomalous and normal samples is equal respectively to 42% and 68% of all test data. The test will perform at one among the IDSs, because all IDSs have the same trained SVM classifier.

## 6.2 Experiments results and discussion

The radial basis function (RBF) is used as the kernel function:

$$F_{RBF} = \exp\left(-\|x_1 - x_2\|/2.\sigma^2\right), \text{where } 1/2.\sigma^2 > 0$$

The accuracy measure is used as performance metric to evaluate our algorithm. We also compute the detection rate, that represents the percentage of correctly detected intrusions, and false positive, that represents the percentage of normal connections that are incorrectly classified as anomalous.

The identification of the most relevant features is an important task, in our scenario we try to determine SVMs-based anomaly detection that achieve high classification accuracy by deleting the useless features. This task is performed by delete one feature at time according to the approach proposed by Sung et al. [18].The increased number of features led a High computational cost in the nodes, for that our aims is to obtain the SVM classifier with less number of features but able to provide high rate of accuracy, in order to save the memory storage and energy consuming in the sensor nodes. The results of the distributed SVMs binary classifier related to the most relevant features with N=18 are summarized in Table 1.

Table 1: The performance evaluation of distributed IDSs based on SVM

| Number of Features | Accuracy (%) | Detection Rate (%) |
|---|---|---|
| 9 | 97.80 | 93.66 |
| 7 | 98.47 | 95.61 |
| 5 | 96.95 | 91.21 |
| 4 | 98.39 | 95.37 |

From Table 1, we find out that, the binary SVM classifier with 7 features outperforms the SVMs that use (9, 5, 4) features, respectively, in terms of accuracy and detection rate. Thus these 7 features represent the most significant features. However, the difference of accuracy between both SVM with 7 and 4 features is small, and due to the resource constraints at sensor nodes, we use SVM with 4 features for anomaly detection engine. These features are:

Src_bytes: Number of bytes sent from source to destination

Dst_bytes: Number of bytes sent from destination to source
Count: Number of connection to same destination host
Srv_diff_host_rate: the percentage of connections to different host

The centralized IDS based on SVM (IDS located in the base station) exhibits high performance for solving a problem of 2-class, but this approach requires all the data to be provided by each sensor. Thus it's consuming much energy. The proposed algorithm is compared to centralized approach in term of classification accuracy by using the selected features (Src_bytes, Dst_bytes, Count, Srv_diff_host_rate). This is illustrated in figure 5, where N is the number of IDSs and sensor nodes for both distributed and centralized approaches respectively.

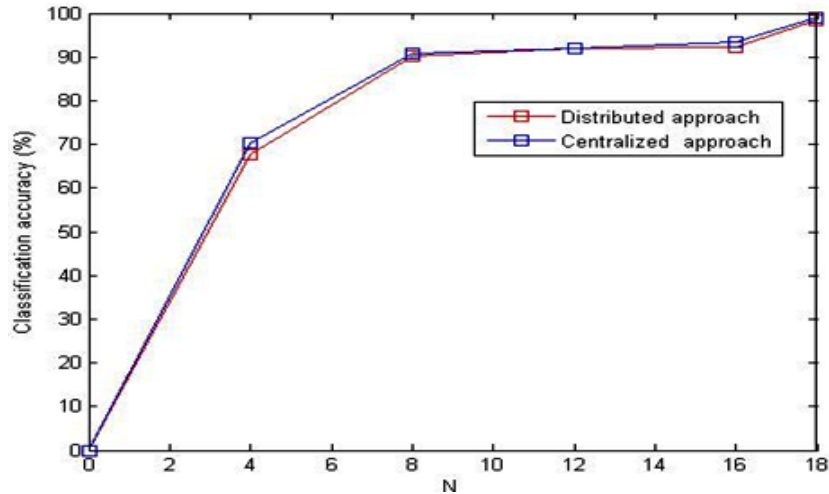

Figure 5.Classification accuracy of SVM for centralized and distributed approaches

As shown in Figure 5, the curves for both approaches coincide almost exactly, and the rate of classification accuracy for centralized and distributed approaches increases when the number of IDSs and sensor nodes increase respectively. Specifically when the number is important (in our case N=18) the rate is close to 100%, 99.07% for centralized approach and 98.39 % for distributed approach. As a result distributed IDSs based on SVM deliver highly-accurate performance, with less training data.

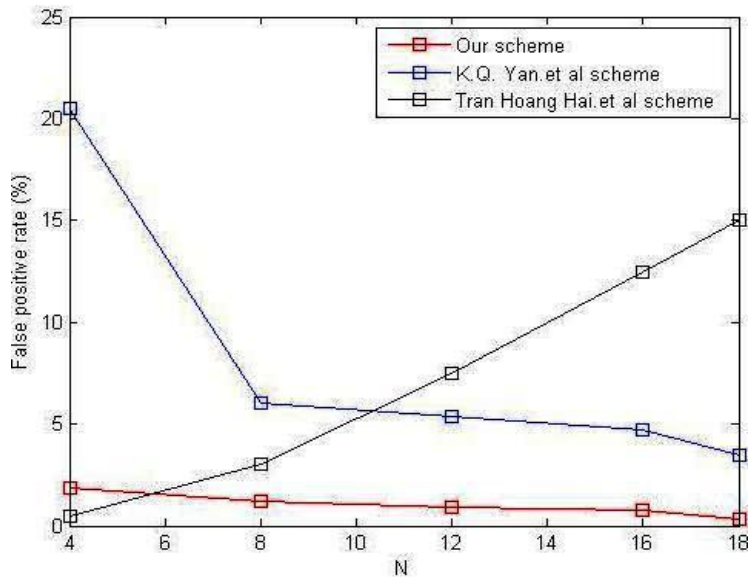

Figure 6.The comparison of the false alarm of different schemes

Increasing the number of IDS nodes for our scheme or sensor nodes (in case of Yan et al. [14] scheme) results in a decrease in the rate of false positive. However, in Hai et al. [13] scheme the number of false positive is important when the number of IDS nodes increase (due to the number of collusion).As shown in Figure 6, our approach exhibits a low false alarm compared to the other schemes, specifically when N=18 ( it's equal to 0.3%). However, the distributed hybrid intrusion detection system proposed in this paper achieves a better effectiveness in terms of a low number of false alarms.

## 7. CONCLUSION AND FUTURE WORK

In this paper, we proposed a distributed hybrid intrusion detection system (HIDSs) for clustered wireless sensor networks. The proposed distributed learning algorithm for the training of SVM in WSN reaches high accuracy for detecting the normal and anomalous behavior (accuracy rate over 98%). Also a combination between the SVM classifier and Signature Based Detection achieve a high detection rate with low false positive rate.

Communication in WSN consumes a high energy, as an example one bit transmitted in WSNs consumes about as much power as executing 800-1000 instructions [24]. The training process is carried out with IDS nodes. These nodes need to compute and transmit only a set of data vector (support vector) between each others, instead of transmitting all captured data to a centralized point, then train a SVM classifier. Thus our approach reduces energy consumption.

In our future work, we will use PSO (particle swarm optimization) to select the relevant features, instead of delete one feature at a time and rank the important one. In the near future, we attempt to implement our approach in sunspot sensor nodes.

**Authors**

**Hichem Sedjelmaci** received his Master degree in Mobile Networks and Service from the University of Tlemcen (Algeria) in 2009. Member of STIC laboratory in the University of Tlemcen, his research interests include the security issue (intrusion detection in particular) of wireless sensor networks and mobile services.

**Mohammed Feham** received his PhD in Engineering in optical and microwave communications from the University of Limoges (France) in 1987, and his PhD in science from the university of Tlemcen (Algeria) in1996. Since 1987, he has been Assistant Professor and Professor of Microwave, Communication Engineering and Telecommunication network. His research interests cover telecommunication systems and mobile networks.